\documentclass[
  aps,             
  pra,             
  twocolumn,       
  superscriptaddress,  
  amsmath,amssymb,
  floatfix,
  nofootinbib
]{revtex4-2}
\usepackage{amssymb}
\usepackage{graphicx}
\usepackage{bm}
\usepackage{bbm}
\usepackage{booktabs}
\usepackage{xcolor}
\usepackage{placeins}   
\usepackage{cleveref}   
\begin{document}

\title{Adaptive AI for Pulse-Level Quantum Control}


\author{Sanjeev Shapkota}
\affiliation{Department of Electrical and Computer Engineering, Southern Methodist University, Dallas, TX 75205, USA}

\author{Yayu Mo}
\affiliation{Department of Electrical and Computer Engineering, Southern Methodist University, Dallas, TX 75205, USA}

%
%
%

\author{Sanjaya Lohani}
\email{slohani@smu.edu}
\affiliation{Department of Electrical and Computer Engineering, Southern Methodist University, Dallas, TX 75205, USA}


\begin{abstract}
\noindent The Control Variational Quantum Eigensolver (ctrl‑VQE) directly optimizes microwave pulses to enable faster and lower‑error quantum‑state preparation, but its continuous control landscape requires efficient search strategies. We demonstrate that a reinforcement‑learning agent based on a deep Q learning network can autonomously discover high‑performance pulse sequences using only system parameters and a reward function. The approach is fully general for superconducting qubit platforms, requires no ansatz, and operates at nanosecond resolution compatible with hardware constraints. As a proof of concept, we apply the method to ground‑state preparation of the Hydrogen molecule on a simulated superconducting device. The agent consistently identifies optimized control sequences that achieve high fidelity and outperform random‑search baselines. These results highlight adaptive learning as a promising hardware‑ready framework for pulse‑level quantum control. 
\end{abstract}


\maketitle


\section{Introduction}
\label{sec:introduction}
As quantum processors scale in complexity, exhibiting dense spectral features, nonlinear interactions, and time-varying hardware imperfections, the task of generating high-fidelity control pulses becomes increasingly challenging for conventional model-based and manually calibrated approaches~\cite{khaneja2005grape, caneva2011crab}. This motivates the integration of adaptive intelligence into the pulse-calibration workflow, where learning algorithms can leverage experimental feedback to iteratively optimize control policies and maintain performance under evolving operating conditions~\cite{baum2021experimental, sivak2022modelfree}. Such adaptability is foundational to the broader promise of quantum computing, where advantages in molecular simulation, materials discovery, and complex optimization depend on the ability to reliably prepare and manipulate quantum states with precision at the lowest level of control. This need is crucial for early quantum applications targeting chemically relevant quantities such as ground-state energies. Variational methods such as VQE have shown promise on small systems~\cite{peruzzo2014variational}, yet their gate-based implementations accumulate error on noisy, shallow-depth hardware~\cite{kandala2017hardware}. Pulse-level variants like ctrl-VQE mitigate this by shaping the microwave drive directly, reducing overhead and aligning the optimization more closely with the device's native physics~\cite{meitei2021gatefree}. In this setting, adaptive learning, particularly reinforcement learning (RL), offers a compelling path forward: learning agents can incorporate feedback, construct pulse sequences dynamically, and maintain performance as hardware conditions evolve~\cite{niu2019universal, bukov2026reinforcement}. We use the Hydrogen molecule as an illustrative example, providing a simple, well-characterized benchmark for evaluating pulse-level adaptive control.

Pulse design has traditionally relied on two major families of classical control methods, Gradient Ascent Pulse Engineering (GRAPE) and Chopped Random Basis (CRAB). GRAPE iteratively refines an initial pulse using gradient information~\cite{khaneja2005grape}, while CRAB simplifies the search by optimizing over a restricted set of basis functions~\cite{caneva2011crab,muller2022decade}. Although both approaches are widely used and effective~\cite{kochkoch2022strategic,leung2017speedup}, they depend on accurate system models to guide optimization and can become trapped in suboptimal solutions when the control landscape is poorly characterized. These limitations naturally motivate model free adaptive intelligent alternatives, leading to reinforcement-learning approaches in which an agent explores pulse sequences directly on the control landscape rather than explicitly relying on the Hamiltonian model, and learns effective strategies from rewards. Additionally, prior work has demonstrated that RL can serve as an effective model-free tool for optimal-control methods~\cite{bukov2018rl,zhang2019when,niu2019universal,fosel2018rl,baum2021experimental,sivak2022modelfree,porotti2019coherent,mackeprang2020rl,an2019deep,dalgaard2020global,chen2014fidelity}. Similarly, In parallel, pulse‑level VQE methods such as ctrl‑VQE and the Pulse‑based Ansatz for VQE (PANSATZ) have shown that bypassing gate decomposition enables more hardware‑native optimization, though these approaches typically rely on gradient‑based updates and, in the case of PANSATZ, restrict the search to a predefined, hardware‑aware pulse set. Motivated by these limitations, we integrate reinforcement learning with ctrl‑VQE to autonomously design pulse sequences that remain compatible across superconducting quantum devices, enabling a more flexible and fully learned approach to pulse‑level variational optimization.

Furthermore, we integrate deep Q learning network (DQN) with RL framework into pulse-level VQE to address the important aspects of ctrl-VQE, which is to identify an effective sequence of pulses within an enormous and highly nonconvex control landscape. Exhaustive brute-force search is infeasible, and gradient-based methods require accurate models, making DQN a natural alternative. In our approach a DQN agent explores pulse sequences, receives rewards based on cost reduction, and gradually learns optimized sequence of pulses that outperform random search while remaining fully model‑free. Note that the agent autonomously constructs pulses, enabling the same learning framework to be applied across different quantum systems updating only the Hamiltonian of interests.

The remainder of the paper is organized as follows. The next section presents the methodological framework, including the mathematical background, the computational model, and the deep Q learning network architecture used to implement reinforcement‑learning‑based pulse optimization. We then report our results, demonstrating the proof‑of‑principle integration of RL with ctrl‑VQE through an illustrative example involving the Hydrogen molecule.

\begin{figure}[h]
    \centering
    \includegraphics[width=\linewidth]{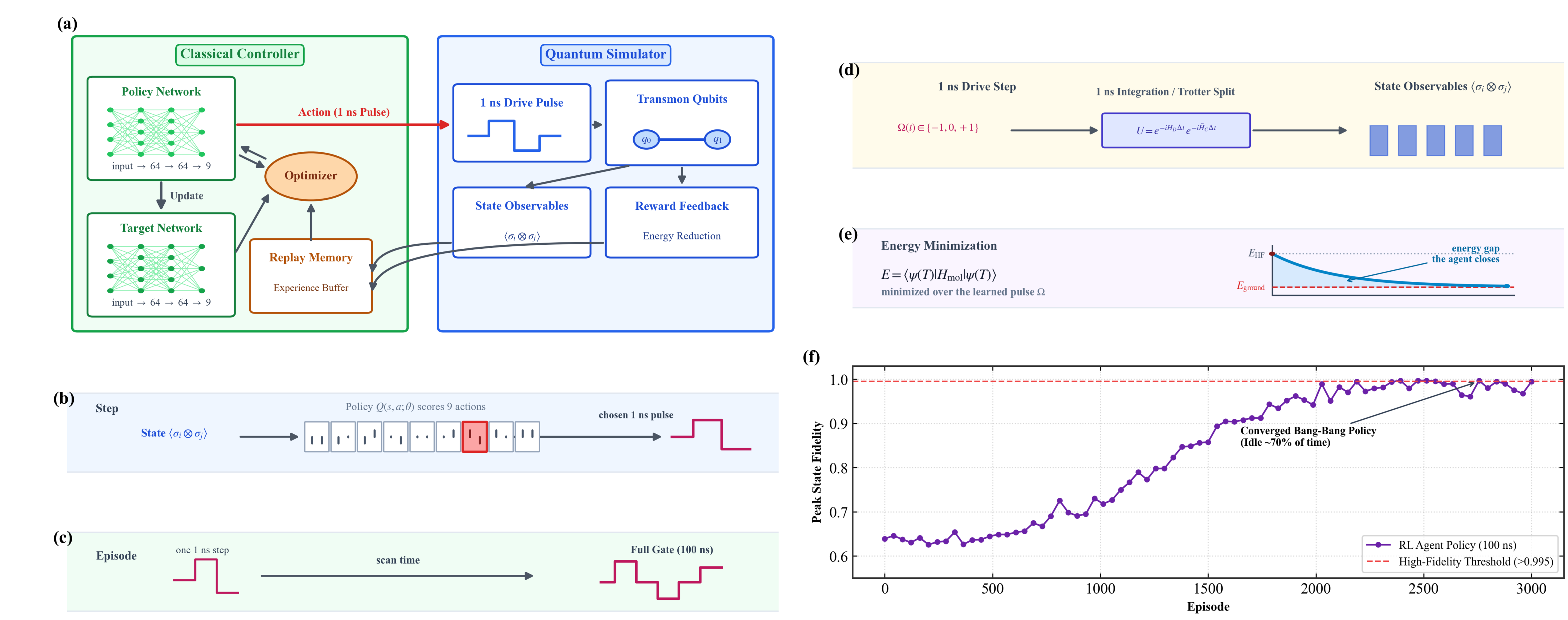}
    \caption{Overview of the method. A reinforcement-learning-agent learns to generate a sequence of control pulses for quantum device.
(a) The complete workflow. The policy network reads the current state and picks a one
nanosecond drive pulse, shown as the red action arrow. This pulse is applied to
the two coupled transmon qubits. The qubits are then measured, and from these
measurements the agent receives the new state and a reward based on how much the
energy dropped. The optimizer uses this reward, together with a target network
and a replay memory, to update the policy network. Both networks have the same
shape: an input layer, two hidden layers of 64 units, and 9 action outputs.
(b) One step. From the nine possible actions, the agent picks one, and
that action sets the drive pulse for that nanosecond. (c) One episode is many such steps joined together to build the full pulse
sequence. (d) Illustrative sketch: Within each step the state evolves under the Hamiltonian, computed
with a first-order Trotter split, and the qubits are measured to give the agent
its next observation. (e) Illustrative sketch: The reward pushes the
energy down from the Hartree-Fock value toward the ground state. (f) The fidelity
rises over training as the agent settles on an optimized policy.}
    \label{fig:overview}
\end{figure}

\section{Methods}

\subsection{Variational quantum eigensolver}
VQE is a hybrid algorithm, a quantum processor prepares a trial state and a
classical optimizer tunes its parameters~\cite{cerezo2021vqa}. It builds on the
variational principle, which says the energy of any trial state can never be lower
than the true ground-state energy,
\begin{equation}
E(\bm{\vartheta}) = \langle\psi(\bm{\vartheta})|H_{\mathrm{mol}}
|\psi(\bm{\vartheta})\rangle \geq E_{\mathrm{ground}}.
\end{equation}
Here $H_{\mathrm{mol}}$ is the qubit Hamiltonian of the molecule and
$E_{\mathrm{ground}}$ is its lowest eigenvalue. The trial state
$\psi(\bm{\vartheta})$ is prepared on the quantum processor and depends on the
parameters $\bm{\vartheta}$.
So by lowering $E(\bm{\vartheta})$ as much as possible over the parameters
$\bm{\vartheta}$, we get closer and closer to the true ground-state energy from
above~\cite{romero2018ucc}.

\subsection{ctrl-VQE: pulse-level control}

The hardware itself is described by the device Hamiltonian~\cite{meitei2021gatefree}
\begin{equation}
H_D = \sum_q \omega_q a_q^{\dagger}a_q
- \sum_q \frac{\delta}{2} a_q^{\dagger}a_q^{\dagger}a_q a_q
+ \sum_{\langle pq\rangle} g_{pq}\, a_p^{\dagger}a_q .
\end{equation}
The first term is the natural oscillation energy of each qubit at frequency
$\omega_q$. The second term is the anharmonicity $\delta$, which makes the energy
gaps between levels unequal so that the drive can address the transition from
$|0\rangle$ to $|1\rangle$ without leaking into higher
levels~\cite{koch2007transmon,krantz2019guide}. The third term is the coupling
$g_{pq}$ between neighboring qubits, which is always on and creates the
entanglement needed to reach the molecular ground state.
The microwave drives signals we control form the control
Hamiltonian~\cite{meitei2021gatefree,asthana2023leakage}
\begin{equation}
H_C = \sum_q \Omega_q(t)\left(e^{i\nu_q t}a_q + e^{-i\nu_q t}a_q^{\dagger}\right),
\label{eq:control}
\end{equation}
In ctrl-VQE we do not use quantum gates. Instead we directly tune the microwave
pulses $\Omega_q(t)$ applied to each qubit $q$~\cite{meitei2021gatefree},
where $\Omega_q(t)$ is the pulse amplitude the agent chooses at every step and
$\nu_q$ is the drive frequency for qubit $q$. The carrier $e^{\pm i\nu_q t}$
delivers the amplitude to the qubit, and the operators $a_q$ and
$a_q^{\dagger}$ rotate the qubit state.
We begin with the Hartree-Fock state $|01\rangle$ and let the pulses act for a
total time $T$. To simplify the dynamics, we work in a rotating frame that
co-rotates with the device, so that only the effect of the pulses needs to be
tracked. In this rotating frame, the state changes like this,
\begin{equation}
|\psi(T)\rangle = e^{-iH_D T}\,\mathcal{T}\exp\!\left(-i\int_0^T \tilde{H}_C(t)\, dt\right)
|01\rangle ,
\end{equation}
where $\tilde{H}_C(t) = e^{iH_D t} H_C(t)\, e^{-iH_D t}$ is the control of
Eq.~\eqref{eq:control} as seen in the rotating frame. The $e^{-iH_D T}$ in front brings the state back to the non-rotating frame at the
final time $T$, just before the energy is measured. The symbol $\mathcal{T}$ means we apply
the small pulse pieces in the right time order, earliest first, because the agent
picks a new pulse every $1$~ns and the order they act in matters.
What we want to make as small as possible is the molecular energy,

\begin{equation}
E= \langle\psi(T)|H_{\mathrm{mol}}|\psi(T)\rangle ,
\end{equation}
which is the average energy of the current state and tells us how close the pulse
has driven the system to the ground state.
We limit the pulse amplitudes to a "bang-bang" form, where each amplitude
can only take one of a few discrete values. For many
optimal-control problems, the fastest solution is exactly ``bang-bang,'' with the
amplitude pushed to its highest or lowest
value~\cite{pontryagin1962optimal,yang2017pmp}. It also turns the problem into a
discrete search, which fits reinforcement learning well.

\subsection{Deep Q learning network}

Q-learning estimates the value of taking an action $a$ in a state
$s$~\cite{sutton2018rl}. This value is the Q-function $Q(s,a)$, the total
reward the agent expects to collect by taking action $a$ in state $s$ and then
making the best choices for the rest of the episode. The agent selects actions with the highest Q-value, and the optimal Q-function
obeys the Bellman equation~\cite{bellman1957dynamic},
\begin{equation}
Q^{*}(s,a) = \mathbb{E}\!\left[r + \gamma \max_{a'} Q^{*}(s',a')\right],
\end{equation}
which says the value of an action is the reward $r$ we get now plus the best
value we can expect later, scaled by a discount factor $\gamma$ that weights
future rewards. Basic Q-learning stores one value for every state-action pair in
a lookup table, but our state space is far too large for that. We therefore
approximate the Q-function with a deep learning network.


We employ a DQN~\cite{mnih2015human} framework in which the optimal action-value function $Q^*(s,a)$ is approximated by a neural network $Q(s,a;\theta)$ parameterized by $\theta$. Two instances of this network are maintained during training: an online Q-network $Q(s,a;\theta)$ and a target Q-network $Q(s,a;\theta_{i-1})$. Both networks share the same architecture but serve distinct roles. The online network is updated at every optimization step and is responsible for estimating the action-value function and selecting actions during interaction with the environment. In contrast, the target network is a slowly updated copy of the online network and is used exclusively to compute stable temporal-difference (TD) targets, thereby mitigating the moving-target instability inherent in bootstrapped value estimation.

To improve sample efficiency and stabilize learning, we employ a replay memory that stores past transitions $(s,a,r,s')$ collected during interaction with the environment. At each training step, we draw a minibatch $B$ of transitions uniformly at random from this buffer, which breaks the correlations between sequential observations and provides a more diverse training distribution. For exploration, we use the standard $\varepsilon$-greedy strategy: with probability $\varepsilon$ the agent selects a random action, and with probability $1-\varepsilon$ it selects the action that maximizes the current Q-value estimate. As indicated in Table \ref{tab:hyperparams}, we begin training with a relatively large $\varepsilon$ and gradually decay it so that the agent transitions from exploration to exploitation as learning progresses.

For each transition $(s,a,r,s')$ in the sampled minibatch, we compute the TD target using the target network parameters $\theta_{i-1}$:
\begin{equation}
    y = r + \gamma \max_{a'} Q(s', a'; \theta_{i-1}),
\end{equation}
where the target network performs both action selection and evaluation. Because the online network $Q(s,a;\theta)$ is not used to select the maximizing action in the TD target, our implementation corresponds to the classical DQN formulation rather than Double DQN.

The TD error for each transition is defined as
\begin{equation}
    \xi = y - Q(s,a;\theta),
\end{equation}
and the online network parameters $\theta$ are optimized by minimizing the Huber loss applied to $\xi$:
\begin{equation}
L(\xi) =
\begin{cases}
\frac{1}{2}\xi^2, & \text{if } |\xi| \leq 1, \\
|\xi| - \frac{1}{2}, & \text{otherwise}.
\end{cases}
\end{equation}
The total loss minimized over the minibatch is the average Huber loss:
\begin{equation}
    L(\theta) = \frac{1}{|B|} \sum_{(s,a,r,s') \in B} L(\xi),
\end{equation}
which corresponds to PyTorch's \texttt{SmoothL1Loss}. Gradients of this loss are computed over the minibatch and applied using the AdamW optimizer.

To ensure stable learning, the target network parameters are updated using a soft-update rule
\begin{equation}
    \theta_{i-1} \leftarrow \tau \theta + (1-\tau)\theta_{i-1},
\end{equation}
with $\tau = 0.005$, ensuring that the target network evolves slowly relative to the online network

\subsection{Experimental setup}
Our test molecule is the Hydrogen dimer, written as a two-qubit problem in the
STO-3G (Slater-Type Orbital approximated by three Gaussians) basis~\cite{hong2022daubechies}. We run it on a simulated transmon device
with fixed frequencies and always-on coupling~\cite{meitei2021gatefree,asthana2023leakage}. When the two atoms sit
$0.75$~\AA\ apart, the exact energy is $-1.1371$~Ha and the Hartree-Fock energy is
$-1.1175$~Ha. The gap between these two numbers is what the agent has to find.

Our earlier study~\cite{shapkota2026rl} used four actions and a $0.1$~ns step. The
agent reached a fidelity of $0.99577$ in about $31$~ns, about three times faster
than the best random search pulse, which reached $0.99449$ in about $88$~ns.

We change two things here. First, each qubit now has three amplitude levels to
choose from, full negative, zero, or full positive, and the policy network picks
one for each qubit at every step. The zero level applies no drive, so it lets the
qubit idle. With two qubits and three levels each, the policy online network chooses
from nine possible actions per step.
Second, we fix each step to $1$~ns so it matches what real waveform generators can
produce~\cite{smith2022pulse}. We try three scanning times, $10$, $40$, and $100$~ns, the duration each pulse
is applied, which makes episodes $10$, $40$, or $100$ steps long. That gives
$9^{10}$, $9^{40}$, or $9^{100}$ possible pulses to choose from. We train for
$3000$ episodes at each scanning time using the same random seed. To judge how close
we get, we mark a $5$~milli-Hartree energy as the reference line.

Furthermore, the agent’s observation consists of the current set of Pauli expectation values ($\langle \sigma_i, \sigma_j \rangle;\, i, \,j = \{I, X, Y, Z\} $), which characterize the quantum state and would be obtained from repeated measurements on real hardware. Based on this observation, the agent selects the drive amplitudes for the next 
$1$~ns control segment from a discrete set of nine possible choices. Appending a segment corresponds to allowing the system to evolve for $1$~ns , after which the agent receives a new state description and continues the process.


The reward is defined at each step. Let $E_t$ be the energy after step $t$ and
$E_\mathrm{min}$ the lowest energy seen so far in training. The reward is
\begin{equation}
r_t \mathrel{=}
\begin{cases}
-10\,(E_t - E_\mathrm{FCI}) + 5 & \text{if } t \mathrel{=} \text{terminal},\ E_t \leq E_\mathrm{min},\\
-10\,(E_t - E_\mathrm{FCI})     & \text{if } t \mathrel{=} \text{terminal},\ E_t > E_\mathrm{min},\\
5                                & \text{if } t \neq \text{terminal},\ E_t \leq E_\mathrm{min},\\
0                                & \text{if } t \neq \text{terminal},\ E_t > E_\mathrm{min}.
\end{cases}
\end{equation}
where $E_\mathrm{FCI}$ is the exact ground-state energy. The energy term
$-10(E_t - E_\mathrm{FCI})$ is given only on the final step and grows as the
final energy drops closer to the exact value, the factor $10$ rescales the small
energy difference into a useful learning signal. The bonus of $5$ is added at any
step that reaches a new lowest energy, after which $E_\mathrm{min}$ is updated to
that value. This bonus encourages the agent to continually surpass its previous best performance.
Each episode corresponds to a full traversal of the scanning window, beginning from the Hartree--Fock state $|01\rangle$ and terminating when the allotted evolution time is exhausted.
Note that the DQN accumulates experience over time. By storing and learning from past transitions, it gradually develops a policy that identifies which pulse adjustments are effective in different situations.

\subsection{Model architecture and baseline random search methods}

\begin{table}[!t]
\caption{All hyperparameters used for training the DQN agent. The same settings
are applied at every scanning time; only the total gate-time budget varies across
experiments.}
\label{tab:hyperparams}
\begin{ruledtabular}
\begin{tabular}{l p{0.62\columnwidth}}
Setting & Value \\
\hline
Network & $16 \to 64 \to 64 \to 9$ (2 hidden layers of 64, 9 actions out) \\
Algorithm & DQN  \\
Optimizer & AdamW (AMSGrad), learning rate $10^{-4}$ \\
Loss & Huber (smooth $L_1$) \\
Gradient clipping & value 1.0 \\
Batch size & 128 \\
Discount factor $\gamma$ & 0.99 \\
Replay buffer capacity & 50{,}000 transitions \\
Target-network update & soft, $\tau = 0.005$ per step \\
Exploration & $\varepsilon$ from 0.9 to 0.05, decay constant $900\times$ scanning time (in ns) \\
Episodes per gate time & 3000 \\
Pulse step width & 1 ns \\
Actions & 9 (bang-bang, includes idle) \\
\end{tabular}
\end{ruledtabular}
\end{table}

We configure our DQN as a fully connected feedforward neural network that maps the agent's observation, the vector of Pauli expectation values describing the current quantum state, to a set of discrete action values. The input layer therefore has dimension equal to the number of measured Pauli operators, 16. We process this input through two hidden layers of 64 neurons each, using ReLU activations to provide sufficient nonlinearity for modeling the control landscape.
The output layer contains nine units, corresponding to the nine available drive-amplitude choices for the next $1\,\text{ns}$ control segment. Each output represents the estimated value $Q(s,a;\theta)$ for selecting that particular pulse configuration. Moreover, the target network $Q(s,a;\theta_{i-1})$ shares the same architecture and initialization as the policy network, but it is updated only through the soft-update rule described above. The policy network is trained using the AdamW optimizer with a learning rate of $10^{-4}$, and all weights use PyTorch's default Kaiming-uniform initialization.
Similarly, the random-search baseline operates under identical conditions, using the same environment, simulator, scanning window, and number of episodes, but selects every action uniformly at random. It contains no network and retains no memory of previous episodes, making it a clean baseline that isolates the contribution of the learned policy.
Note that we simulate the evolution of the quantum system on a classical computer using Trotterization. This approach mirrors the sequence of operations a quantum processor would carry out by applying the control pulse, allowing the system to evolve under the Hamiltonian, and then evaluating the resulting state.
The hyperparameter set used in this paper is shown in Table~\ref{tab:hyperparams}.

\section{Results and Discussion}
\label{sec:results}
\subsection{Cumulative sub-Hartree--Fock discovery}
\begin{figure}[t]
\centering
\includegraphics[width=\linewidth]{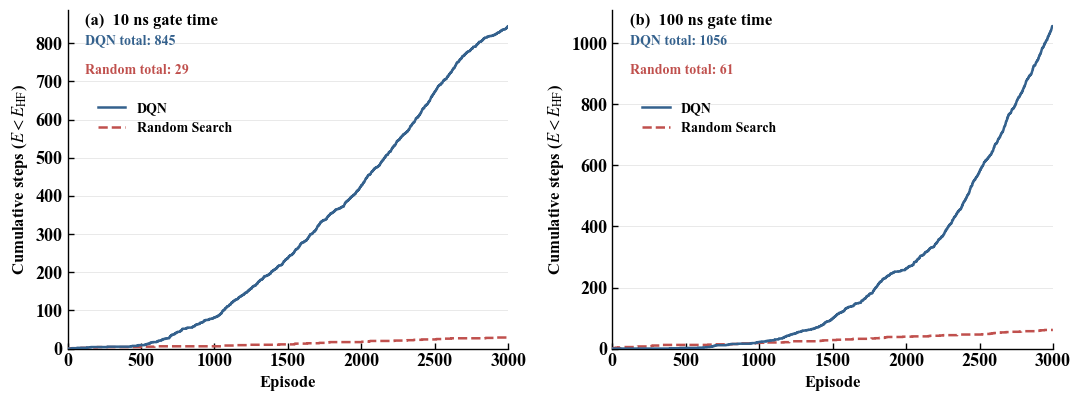}
\caption{Cumulative number of sub-Hartree--Fock steps ($E < E_{\mathrm{HF}}$)
versus training episode for the DQN policy (blue) and the random-search baseline
(red), at representative scanning times of $10$~ns (left) and $100$~ns (right).
The DQN curve accelerates with training while random search remains slow and
nearly linear.}
\label{fig:cumulative}
\end{figure}

We first measure how often each method finds energies below the Hartree--Fock
baseline, $E < E_{\mathrm{HF}}$. Finding such a state means the agent has reached
the quantum-correlated regime that matters for estimating the ground state. The
total number of findings over $3000$ training episodes is shown in
Fig.~\ref{fig:cumulative}, with results displayed for both a shorter (left) and a
longer (right) scanning time. The DQN curve starts almost flat and then bends
sharply upward as training goes on, a sign of learning, while the random-search
curve grows slowly and steadily throughout, with no sign of the speed-up that
comes from learning.
At $10$~ns the DQN agent finds $845$ sub-Hartree--Fock states over training while
random search finds only $29$---about $29$ times more. At $100$~ns DQN finds
$1056$ against $61$ for random search---about $17$ times more. The very large
ratio at $10$~ns comes from the short search horizon of only ten pulse steps,
which the agent learns quickly. At longer scanning times the number of steps $N$
increases, and the number of possible pulse sequences grows as $9^{N}$. Random
search therefore struggles as $N$ grows, yet DQN still finds far more useful
states. The largest total, $1056$, occurs at $100$~ns, showing that the agent may
become more effective as we increase the scanning window.

\subsection{Success rate over training}

\begin{figure}[!t] 
\centering
\includegraphics[width=\linewidth]{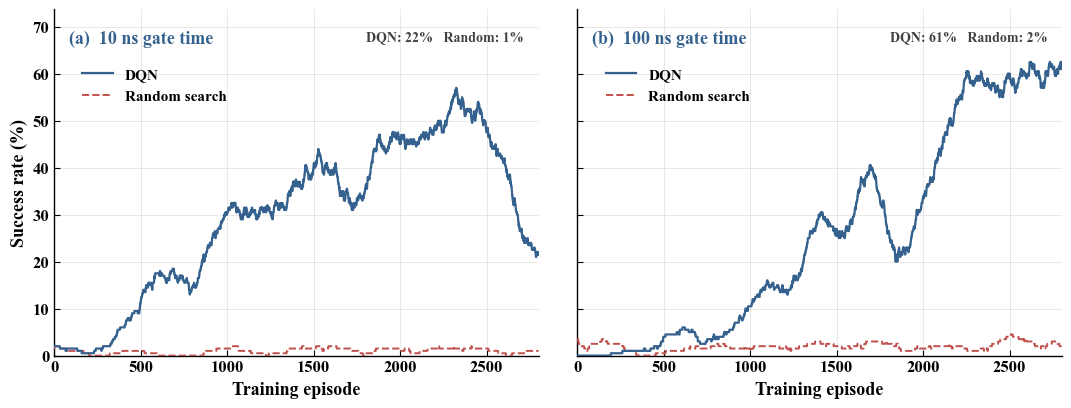}
\caption{Success rate (fraction of episodes with $E < E_{\mathrm{HF}}$, 200-episode moving average) versus training episode for the Hydrogen dimer at $0.75$~\AA\ (STO-3G), at representative scanning times of (a) $10$~ns and (b) $100$~ns. The DQN policy (solid blue) improves steadily with training while the random-search baseline (dashed red) stays flat.}
\label{fig:success}
\end{figure}

We also track the success rate, defined as the fraction of episodes within a
sliding window of $200$ episodes in which the agent finds at least one energy
below the Hartree-Fock baseline. Using a moving window of $200$ episodes rather
than single episodes smooths out run-to-run noise and makes the training trend
easier to see. Figure~\ref{fig:success} shows how this success rate evolves over
training.

At $10$~ns the DQN success rate rises to a peak of roughly $57\%$ near episode
$2350$, then declines as indicated in the Figure \ref{fig:success}(a). The late-stage decline is unexpected, but it may
indicate a trade-off between extended training and maintaining peak accuracy for a shorter scanning window. At $100$~ns
it climbs more steadily, reaching about $61\%$ by the end. In both cases the
random-search baseline stays flat near $1$--$2\%$. Random search starts each
episode fresh with no memory of past runs, so its success rate does not improve
as training continues. The rising DQN curve next to the flat baseline is a clear
sign the agent is learning. At $100$~ns, where the gap is largest, DQN finds a
useful energy about $30$ times more often than random search by the end of
training.

\begin{figure}[!t]
\centering
\includegraphics[width=\linewidth]{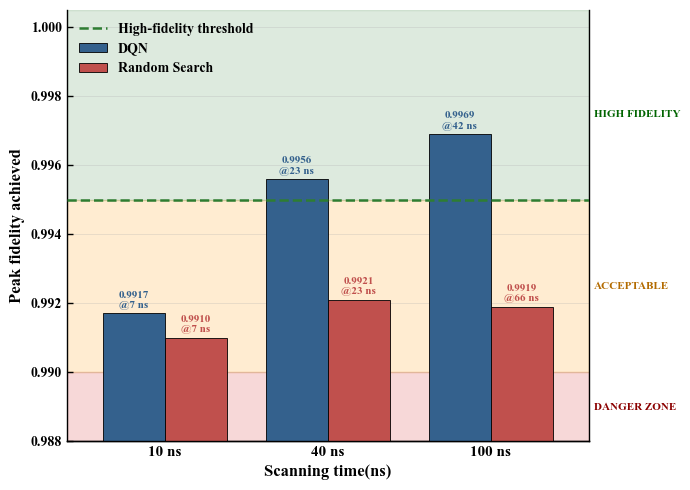}
\caption{Peak state fidelity achieved by the DQN policy (blue) and random search
(red) at the total scanning times of $10$, $40$, and $100$~ns. Annotations give the fidelity and
the pulse duration at which it is reached. The DQN policy attains higher peak
fidelity at every scanning time and is the only method to enter the high-fidelity
region above $0.995$.}
\label{fig:peakfid}
\end{figure}
\subsection{Fidelity and decoherence}

We next look at the best (peak) fidelity each method reaches and how long it takes
to get there. The peak fidelity at scanning times of
$10$, $40$, and $100$~ns is shown in Figure~\ref{fig:peakfid}. DQN reaches a higher peak fidelity at every scanning time,
going from $0.9917$ at $10$~ns to $0.9956$ at $40$~ns and $0.9969$ at $100$~ns,
while random search stays around $0.991$--$0.992$. The gap between them widens as
the scanning time grows, and only DQN climbs into the high-fidelity region above
$0.995$.

At total scanning time of $100\,\text{ns}$, the DQN reaches its highest fidelity much
earlier than the random-search baseline. As shown in Figure~\ref{fig:peakfid}, the
DQN attains its peak at approximately $42\,\text{ns}$, whereas random search
requires about $66\,\text{ns}$. This difference of roughly $24\,\text{ns}$ is
significant because the quantum state gradually loses coherence as the time passes. Reaching the peak sooner reduces the likelihood that the state
degrades before measurement. For this reason, the $100\,\text{ns}$ setting
provides the most favorable balance in our tests, offering both the highest
fidelity and the largest reduction in evolution time.

\subsection{Energy accuracy}

\begin{figure}[!t]
\centering
\includegraphics[width=\linewidth]{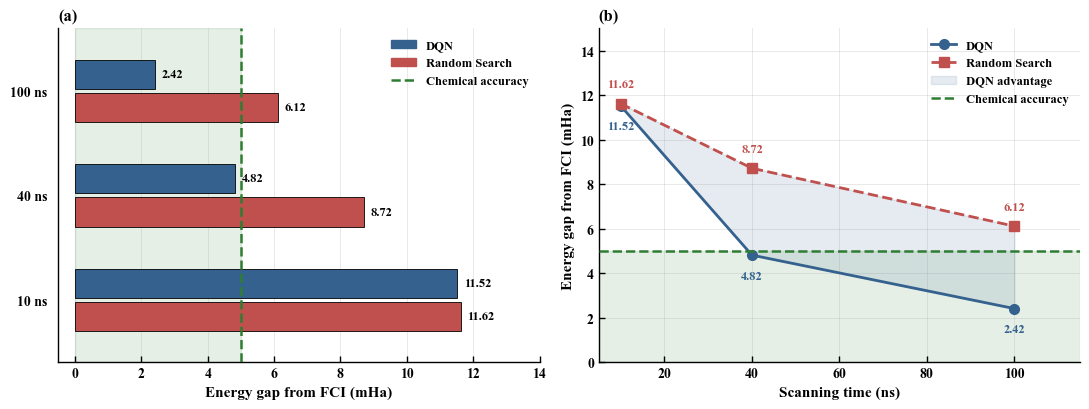}
\caption{Absolute energy gap from the FCI reference for the Hydrogen dimer at
$0.75$~\AA\ (STO-3G). (a) Gap at each scanning time for the DQN policy (blue) and
random search (red); the dashed line and shaded region mark the $5$~mHa accuracy
reference. (b) Energy-gap trend versus scanning time, with the shaded region
indicating the DQN advantage. The DQN policy is consistently closer to FCI and the
advantage widens with scanning time.}
\label{fig:egap}
\end{figure}

What we ultimately care about is how close each method gets to the exact ground
state energy. The energy gap from FCI at each pulse
time, with a $ 5$~milli-Hartree accuracy reference drawn as a dashed line and shaded zone is shown in Figure~\ref{fig:egap}.

Panel~(a) shows the result directly, at every scanning time the DQN bar is shorter
than the random-search bar, so DQN always reaches a lower energy. The DQN gaps are
$11.52$, $4.82$, and $2.42$~mHa at $10$, $40$, and $100$~ns, against $11.62$,
$8.72$, and $6.12$~mHa for random search. At $10$~ns, with only ten pulse steps,
the two methods are almost tied (about $1\%$ apart), but the gap grows quickly with
scanning time, DQN's energy error is $45\%$ lower at $40$~ns and $60\%$ lower at
$100$~ns. Compared to the $5$~mHa reference, DQN enters the accuracy zone at
$40$~ns and stays inside it at $100$~ns, while random search never reaches it.
Panel~(b) shows the same data as a trend line, both curves drop as the scanning time
grows, but the DQN curve drops faster and the gap between them widens. This means
DQN keeps pulling further ahead as the pulse gets longer and the $9^{N}$ space of
possible sequences grows.

\subsection{Structure of the learned policy and control pulses}

\begin{figure}[!t]
\centering
\includegraphics[width=\linewidth]{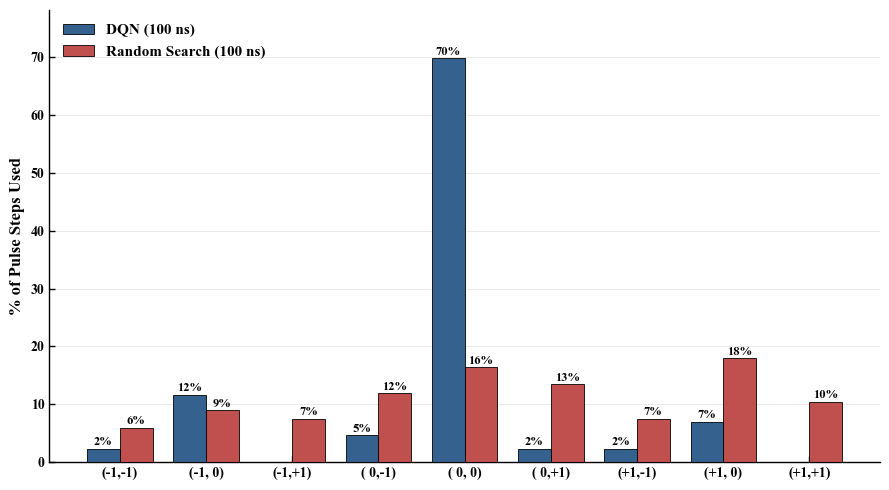}
\caption{Action-usage distribution at the $100$~ns scanning time for the DQN policy
(blue) and random search (red), expressed as the percentage of pulse steps
assigned to each of the nine action-spaces $(\Delta_1,\Delta_2) \in
\{-1,0,+1\}^2$. For each method, the analysis covers the best sequence up to the
step at which it reaches its peak fidelity ($43$ steps for DQN, $67$ for random
search as shown in Figure~\ref{fig:peakfid}). The DQN policy concentrates heavily on the
idle action $(0,0)$, whereas random search spreads its usage nearly uniformly.}
\label{fig:action}
\end{figure}

We explore which actions the training agent actually uses.
The comparison between the DQN and random search action-space  at each evolution for
$100$~ns is shown in Figure~\ref{fig:action}.

Random search uses all nine actions with roughly equal frequency, each between
about $6\%$ and $18\%$, close to the $1/9 \approx 11.1\%$ expected under uniform
random choice. DQN behaves very differently. It spends about $70\%$ of its steps
on the idle action $(0,0)$, uses most of the remainder on single-qubit drives,
and almost never drives both qubits at once; two of the simultaneous two-qubit
actions are never used at all. In effect, the agent learns to mostly idle and let
the system evolve on its own, applying short, precise pulses only at select
moments. This pattern, long intervals of free evolution punctuated by occasional
pulses, resembles the bang-bang strategy that optimal-control theory identifies
as the most time-efficient way to drive a quantum system to a target
state~\cite{pontryagin1962optimal,yang2017pmp}. This result suggests that the agent may be learning a control strategy that aligns with the underlying physics of the system, based solely on the reward signals it receives.

\begin{figure}[!t]
    \centering
    \includegraphics[width=\linewidth]{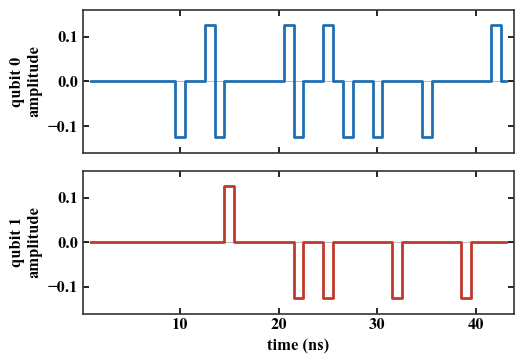}
    \caption{Learned control pulses for the two qubits, showing the best pulse found
by the DQN policy, which reaches a peak fidelity of $0.9969$ at $42$~ns as shown in Figure~\ref{fig:peakfid}.
The top panel is qubit~$0$ and the bottom panel is qubit~$1$. At each
one-nanosecond step, the agent either drives a qubit at full strength (positive
or negative) or leaves it idle, with no intermediate amplitudes. Both qubits stay
idle for most of the time and fire only brief pulses at selected moments, and the
two qubits are driven at different times.}
    \label{fig:pulse}
\end{figure}


We now look at the actual pulses the training agent finds at the $100$~ns scanning time.
The drive on each of the two qubits over the full
scanning time is shown in Figure~\ref{fig:pulse}. At every one nanosecond step the agent either drives a qubit at full
strength, up or down, or leaves it idle, with nothing in between. As described in the previous paragraph and indicated in Figure~\ref{fig:action}, both qubits stay
idle for most of times and fire only short pulses at a few selected moments, and
the two qubits are driven at different times.

\section{Discussion and Conclusion}

We explored whether a learning-based approach can assist in
constructing pulse sequences for state preparation. The agent receives only the
system description and a numerical reward, and it attempts to improve its
performance over repeated trials. Across the settings we examined, the agent
performed better than a random-search baseline, particularly in cases where the
number of possible pulse choices is large. These results suggest that learning
methods may be useful when the search space becomes too extensive for simple
trial-and-error strategies.

The pulse sequences produced by the agent also exhibit a pattern in how control
is applied. Although the agent is not provided with any control-theoretic
guidance, it tends to trigger pulses at selected moments while
remaining idle otherwise. This resembles strategies that have been studied in
theoretical work. While we do not draw firm conclusions from this observation,
it indicates that learning-based approaches may be capable of identifying
structured control behaviors.

We also observed that the agent's performance is limited by the discrete set of
drive levels available during training. With only three levels per qubit, the
agent cannot reduce the energy beyond a certain point. This appears to be a
constraint of the action space rather than the learning method itself. Allowing
finer or continuous control options may enable further improvement.

Although Hydrogen molecule is used as a test case here, the approach does not rely on
features specific to that system. The method depends only on the ability to
simulate the dynamics and evaluate a reward, suggesting that it could be applied
to other systems with similar requirements. The pulse durations used in training
match the time resolution of current hardware, which may make it possible to
transfer learned pulses to real devices.

Our device model incorporates realistic parameters, but does not include all
sources of noise present in physical systems, such as decoherence and control
imperfections. Future work may expand this approach to larger molecular systems and materials, train in more realistic noise models, and evaluate directly on hardware. These directions aim to further assess how learning-based techniques can support the development of control strategies that align with the capabilities of quantum devices.

\section*{Acknowledgments}
The authors thank Chenxu Liu, Yanzhu Chen, Brian T. Kirby, and Thomas A. Searles for valuable discussions. We acknowledge support from the SMU startup fund, and Sanjaya Lohani also acknowledges support from the Sam Taylor Fellowship Program. Computational resources for this work were provided in part by SMU’s O’Donnell Data Science and Research Computing Institute.

\bibliographystyle{apsrev4-2}
\bibliography{references}   

@article{cerezo2021vqa,
  author  = {Cerezo, M. and Arrasmith, A. and Babbush, R. and Benjamin, S. C. and Endo, S. and Fujii, K. and McClean, J. R. and Mitarai, K. and Yuan, X. and Cincio, L. and Coles, P. J.},
  title   = {Variational quantum algorithms},
  journal = {Nature Reviews Physics},
  volume  = {3},
  number  = {9},
  pages   = {625--644},
  year    = {2021}
}

@article{romero2018ucc,
  author  = {Romero, J. and Babbush, R. and McClean, J. R. and Hempel, C. and Love, P. J. and Aspuru-Guzik, A.},
  title   = {Strategies for quantum computing molecular energies using the unitary coupled cluster ansatz},
  journal = {Quantum Science and Technology},
  volume  = {4},
  number  = {1},
  pages   = {014008},
  year    = {2018}
}

@article{kandala2017hardware,
  author  = {Kandala, Abhinav and Mezzacapo, Antonio and Temme, Kristan and Takita, Maika and Brink, Markus and Chow, Jerry M. and Gambetta, Jay M.},
  title   = {Hardware-efficient variational quantum eigensolver for small molecules and quantum magnets},
  journal = {Nature},
  volume  = {549},
  pages   = {242--246},
  year    = {2017}
}

@article{meitei2021gatefree,
  author  = {Meitei, O. R. and Gard, B. T. and Barron, G. S. and Pappas, D. P. and Economou, S. E. and Barnes, E. and Mayhall, N. J.},
  title   = {Gate-free state preparation for fast variational quantum eigensolver simulations},
  journal = {npj Quantum Information},
  volume  = {7},
  number  = {1},
  pages   = {155},
  year    = {2021}
}

@article{khaneja2005grape,
  author  = {Khaneja, Navin and Reiss, Timo and Kehlet, Cindie and Schulte-Herbr{\"u}ggen, Thomas and Glaser, Steffen J.},
  title   = {Optimal control of coupled spin dynamics: design of {NMR} pulse sequences by gradient ascent algorithms},
  journal = {Journal of Magnetic Resonance},
  volume  = {172},
  pages   = {296--305},
  year    = {2005}
}

@article{caneva2011crab,
  author  = {Caneva, Tommaso and Calarco, Tommaso and Montangero, Simone},
  title   = {Chopped random-basis quantum optimization},
  journal = {Physical Review A},
  volume  = {84},
  pages   = {022326},
  year    = {2011}
}

@article{muller2022decade,
  author  = {M{\"u}ller, Matthias M. and Said, Ressa S. and Jelezko, Fedor and Calarco, Tommaso and Montangero, Simone},
  title   = {One decade of quantum optimal control in the chopped random basis},
  journal = {Reports on Progress in Physics},
  volume  = {85},
  pages   = {076001},
  year    = {2022}
}

@article{kochkoch2022strategic,
  author  = {Koch, Christiane P. and Boscain, Ugo and Calarco, Tommaso and Dirr, Gunther and Filipp, Stefan and Glaser, Steffen J. and Kosloff, Ronnie and Montangero, Simone and Schulte-Herbr{\"u}ggen, Thomas and Sugny, Dominique and Wilhelm, Frank K.},
  title   = {Quantum optimal control in quantum technologies. {Strategic} report on current status, visions and goals for research in {Europe}},
  journal = {EPJ Quantum Technology},
  volume  = {9},
  pages   = {19},
  year    = {2022}
}

@article{leung2017speedup,
  author  = {Leung, Nelson and Abdelhafez, Mohamed and Koch, Jens and Schuster, David},
  title   = {Speedup for quantum optimal control from automatic differentiation based on graphics processing units},
  journal = {Physical Review A},
  volume  = {95},
  pages   = {042318},
  year    = {2017}
}

@book{pontryagin1962optimal,
  author    = {Pontryagin, L. S. and Boltyanskii, V. G. and Gamkrelidze, R. V. and Mishchenko, E. F.},
  title     = {The Mathematical Theory of Optimal Processes},
  publisher = {Interscience Publishers},
  address   = {New York},
  year      = {1962}
}

@article{yang2017pmp,
  author  = {Yang, Z. C. and Rahmani, A. and Shabani, A. and Neven, H. and Chamon, C.},
  title   = {Optimizing variational quantum algorithms using {Pontryagin's} minimum principle},
  journal = {Physical Review X},
  volume  = {7},
  number  = {2},
  pages   = {021027},
  year    = {2017}
}

@article{bukov2018rl,
  author  = {Bukov, Marin and Day, Alexandre G. R. and Sels, Dries and Weinberg, Phillip and Polkovnikov, Anatoli and Mehta, Pankaj},
  title   = {Reinforcement learning in different phases of quantum control},
  journal = {Physical Review X},
  volume  = {8},
  pages   = {031086},
  year    = {2018}
}

@article{zhang2019when,
  author  = {Zhang, Xiao-Ming and Wei, Zezhu and Asad, Raza and Yang, Xu-Chen and Wang, Xin},
  title   = {When does reinforcement learning stand out in quantum control? {A} comparative study on state preparation},
  journal = {npj Quantum Information},
  volume  = {5},
  pages   = {85},
  year    = {2019}
}

@article{niu2019universal,
  author  = {Niu, Murphy Yuezhen and Boixo, Sergio and Smelyanskiy, Vadim N. and Neven, Hartmut},
  title   = {Universal quantum control through deep reinforcement learning},
  journal = {npj Quantum Information},
  volume  = {5},
  pages   = {33},
  year    = {2019}
}

@article{fosel2018rl,
  author  = {F{\"o}sel, Thomas and Tighineanu, Petru and Weiss, Talitha and Marquardt, Florian},
  title   = {Reinforcement learning with neural networks for quantum feedback},
  journal = {Physical Review X},
  volume  = {8},
  pages   = {031084},
  year    = {2018}
}

@article{baum2021experimental,
  author  = {Baum, Yuval and Amico, Mirko and Howell, Sean and Hush, Michael and Liuzzi, Maggie and Mundada, Pranav and Merkh, Thomas and Carvalho, Andre R.R. and Biercuk, Michael J.},
  title   = {Experimental deep reinforcement learning for error-robust gate-set design on a superconducting quantum computer},
  journal = {PRX Quantum},
  volume  = {2},
  pages   = {040324},
  year    = {2021}
}

@article{sivak2022modelfree,
  author  = {Sivak, V. V. and Eickbusch, A. and Liu, H. and Royer, B. and Tsioutsios, I. and Devoret, M. H.},
  title   = {Model-free quantum control with reinforcement learning},
  journal = {Physical Review X},
  volume  = {12},
  pages   = {011059},
  year    = {2022}
}

@article{porotti2019coherent,
  author  = {Porotti, Riccardo and Tamascelli, Dario and Restelli, Marcello and Prati, Enrico},
  title   = {Coherent transport of quantum states by deep reinforcement learning},
  journal = {Communications Physics},
  volume  = {2},
  pages   = {61},
  year    = {2019}
}

@article{mackeprang2020rl,
  author  = {Mackeprang, Jacob and Dasari, Durga B. Rao and Wrachtrup, J{\"o}rg},
  title   = {A reinforcement learning approach for quantum state engineering},
  journal = {Quantum Machine Intelligence},
  volume  = {2},
  pages   = {5},
  year    = {2020}
}

@article{an2019deep,
  author  = {An, Zheng and Zhou, D. L.},
  title   = {Deep reinforcement learning for quantum gate control},
  journal = {EPL (Europhysics Letters)},
  volume  = {126},
  pages   = {60002},
  year    = {2019}
}

@article{dalgaard2020global,
  author  = {Dalgaard, Mogens and Motzoi, Felix and S{\o}rensen, Jens Jakob and Sherson, Jacob},
  title   = {Global optimization of quantum dynamics with {AlphaZero} deep exploration},
  journal = {npj Quantum Information},
  volume  = {6},
  pages   = {6},
  year    = {2020}
}

@article{chen2014fidelity,
  author  = {Chen, Chunlin and Dong, Daoyi and Li, Han-Xiong and Chu, Jian and Tarn, Tzyh-Jong},
  title   = {Fidelity-based probabilistic {Q}-learning for control of quantum systems},
  journal = {IEEE Transactions on Neural Networks and Learning Systems},
  volume  = {25},
  pages   = {920--933},
  year    = {2014}
}

@inproceedings{shapkota2026rl,
  author    = {Shapkota, S. and Mo, Y. and Liu, C. and Chen, Y. and Kirby, B. T. and Searles, T. A. and Lohani, S.},
  title     = {Reinforcement learning for control variational quantum algorithm},
  booktitle = {Proc. IEEE Dallas Circuits and Systems Conference (DCAS)},
  address   = {Dallas, TX},
  year      = {2026}
}

@article{smith2022pulse,
  author  = {Smith, K. N. and Ravi, G. S. and Alexander, T. and Bronn, N. T. and Carvalho, A. R. and Cervera-Lierta, A. and Chong, F. T. and Chow, J. M. and Cubeddu, M. and Hashim, A. and others},
  title   = {Programming physical quantum systems with pulse-level control},
  journal = {Frontiers in Physics},
  volume  = {10},
  pages   = {900099},
  year    = {2022}
}

@article{koch2007transmon,
  author  = {Koch, Jens and Yu, Terri M. and Gambetta, Jay and Houck, A. A. and Schuster, D. I. and Majer, J. and Blais, Alexandre and Devoret, M. H. and Girvin, S. M. and Schoelkopf, R. J.},
  title   = {Charge-insensitive qubit design derived from the {Cooper} pair box},
  journal = {Physical Review A},
  volume  = {76},
  pages   = {042319},
  year    = {2007}
}

@article{krantz2019guide,
  author  = {Krantz, Philip and Kjaergaard, Morten and Yan, Fei and Orlando, Terry P. and Gustavsson, Simon and Oliver, William D.},
  title   = {A quantum engineer's guide to superconducting qubits},
  journal = {Applied Physics Reviews},
  volume  = {6},
  pages   = {021318},
  year    = {2019}
}

@article{asthana2023leakage,
  author  = {Asthana, A. and Liu, C. and Meitei, O. R. and Economou, S. E. and Barnes, E. and Mayhall, N. J.},
  title   = {Leakage reduces device coherence demands for pulse-level molecular simulations},
  journal = {Physical Review Applied},
  volume  = {19},
  pages   = {064071},
  year    = {2023}
}

@book{sutton2018rl,
  author    = {Sutton, Richard S. and Barto, Andrew G.},
  title     = {Reinforcement Learning: An Introduction},
  edition   = {2nd},
  publisher = {MIT Press},
  address   = {Cambridge, MA},
  year      = {2018}
}

@article{mnih2015human,
  author  = {Mnih, Volodymyr and Kavukcuoglu, Koray and Silver, David and Rusu, Andrei A. and Veness, Joel and Bellemare, Marc G. and Graves, Alex and Riedmiller, Martin and Fidjeland, Andreas K. and Ostrovski, Georg and others},
  title   = {Human-level control through deep reinforcement learning},
  journal = {Nature},
  volume  = {518},
  pages   = {529--533},
  year    = {2015}
}

@article{hong2022daubechies,
  author  = {Hong, C.-L. and Tsai, T. and Chou, J.-P. and Chen, P.-J. and Tsai, P.-K. and Chen, Y.-C. and Kuo, E.-J. and Srolovitz, D. and Hu, A. and Cheng, Y.-C. and others},
  title   = {Accurate and efficient quantum computations of molecular properties using {Daubechies} wavelet molecular orbitals},
  journal = {PRX Quantum},
  volume  = {3},
  number  = {2},
  pages   = {020360},
  year    = {2022}
}

@book{bellman1957dynamic,
  title={Dynamic Programming},
  author={Bellman, Richard},
  year={1957},
  publisher={Princeton University Press},
  address={Princeton, NJ}
}

@article{peruzzo2014variational,
  title={A variational eigenvalue solver on a photonic quantum processor},
  author={Peruzzo, Alberto and McClean, Jarrod and Shadbolt, Peter and Yung, Man-Hong and Zhou, Xiao-Qi and Love, Peter J. and Aspuru-Guzik, Al{\'a}n and O'Brien, Jeremy L.},
  journal={Nature Communications},
  volume={5},
  pages={4213},
  year={2014},
  publisher={Nature Publishing Group},
  doi={10.1038/ncomms5213}
}

@article{bukov2026reinforcement,
  title={Reinforcement Learning for Quantum Technology},
  author={Bukov, Marin and Marquardt, Florian},
  journal={arXiv preprint arXiv:2601.18953},
  year={2026}
}

\end{document}